# RESIDUAL-BASED LOCALIZATION AND QUANTIFICATION OF PEAKS IN X-RAY DIFFRACTOGRAMS[1]

By P. L. Davies, U. Gather, M. Meise, D. Mergel and T. Mildenberger

*University Duisburg–Essen and Technische Universität Dortmund*

We consider data consisting of photon counts of diffracted x-ray radiation as a function of the angle of diffraction. The problem is to determine the positions, powers and shapes of the relevant peaks. An additional difficulty is that the power of the peaks is to be measured from a baseline which itself must be identified. Most methods of de-noising data of this kind do not explicitly take into account the modality of the final estimate. The residual-based procedure we propose uses the so-called taut string method, which minimizes the number of peaks subject to a tube constraint on the integrated data. The baseline is identified by combining the result of the taut string with an estimate of the first derivative of the baseline obtained using a weighted smoothing spline. Finally, each individual peak is expressed as the finite sum of kernels chosen from a parametric family.

**1. Introduction.** In the analysis of the morphology of thin films, x-ray diffraction is an indispensable tool [Birkholz (2006)]: the intensity of diffracted x-rays yields important information about the crystalline structure of the material under consideration. The experimental data are usually obtained in the form of a diffractogram: photon counts of x-ray radiation are measured as a function of the angle of diffraction $2\theta$.

A typical diffractogram, as shown in Figure 1, exhibits peaks as well as a slowly varying baseline. The physically relevant information is contained in the location, shape and size of the peaks and their decomposition into a sum of one or more possibly overlapping components represented by kernels. Often thin film diffractograms are analyzed using ad-hoc methods where

Received November 2007; revised May 2008.

[1]Supported in part by Sonderforschungsbereich 475, Technische Universität Dortmund. Also supported by the Collaborative Research Center “Reduction of Complexity in Multivariate Data Structures” (SFB 475) funded by the German Research Foundation (DFG).

*Key words and phrases.* Nonparametric regression, confidence regions, peak detection, x-ray diffractometry, thin film physics.

denoising, removal of the baseline and fitting of the peaks are performed manually. Apart from being inconvenient, this often requires knowledge of possible peak positions.

In this article we suggest a new flexible automatic procedure for the analysis of thin film diffractograms. Our aim is to separate the signal of interest from the noise. More specifically, we aim at a decomposition of the form

$$\text{Data} = \text{Baseline} + \text{Peaks} + \text{Noise}. \tag{1}$$

Our fully automatic five-step procedure (see Section 2 below) removes the baseline and determines the number, positions, powers and shapes of the relevant peaks and their components. It can be applied when little or no prior knowledge of approximate peak positions is available, as is often the case in the analysis of the morphology of thin films. Throughout all stages of the procedure, we employ the following principle: among all models we choose the simplest one which is consistent with the data. That is, "simple" models are favored over "complex" models, but the definition of "simplicity" or, equivalently, "complexity" depends on the particular problem to be solved. We use three different definitions of complexity, namely,

- the number of peaks,
- the value of $\int g^{(2)}(\theta)^2\,d\theta$ as a measure of roughness of the function $g$,
- the number of components or kernels in the representation of each individual peak.

More formally, we first construct an approximation or confidence region (Section 3) using special multiscale conditions for the residuals. This specifies the set of functions consistent with the data. Within this class we then choose a model with minimum complexity [cf. Davies, Kovac and Meise (2008)].

To carry out this program, we make use of recent advances in nonparametric regression and denoising techniques, in particular, the taut string method of Davies and Kovac (2001) and the weighted smoothing splines procedure of Davies and Meise (2008). The taut string method reliably identifies the local extremes of the regression function and it is used to provide initial estimates of the "Peaks" component of (1). Weighted smoothing splines are then used in conjunction with the known positions of the peaks to provide a smooth estimate of the "Baseline" component of (1). Finally, we fit sums of Pearson Type VII curves to the identified peak intervals in order to decompose the peaks into their components and estimate the physically relevant parameters. What remains is the "Noise" component of (1).

We note that the application of the proposed method is not limited to thin film x-ray diffractograms. With little or no modification the procedure could also be applied to other types of diffractograms, for example, of powders or partly crystalline fibers of various materials. A wide range of spectroscopic methods yield data of a similar nature and require the unambiguous and automated identification of the position and width of relatively sharp peaks. Other applications could, for example, include the analysis of Raman-, FTIR- or NMR spectra and mass spectrometry data.

The paper is organized as follows. Section 2 gives some physical background and a description of the data sets as well as a short outline of our method. In Section 3 we introduce the statistical principles on which our procedure is based. Section 4 contains a short description of the taut string method and Section 5 introduces some modifications to accomodate for heteroskedastic noise. Section 6 describes the weighted smoothing splines procedure. Section 7 is devoted to the identification of the baseline and Section 8 to the identification and decomposition of the peaks. Section 9 gives a physical interpretation of the results. Finally, Section 10 contains a short discussion of the complete procedure.

**2. Diffractograms.** X-ray diffraction is an important tool in various fields, including the analysis of crystalline materials, the identification of the molecular structure of proteins and, more recently, also as means to investigate the morphology of thin films. When thin films are prepared on glass substrates they are usually polycrystalline and may even contain different crystalline phases. The experimental data are usually obtained in the form of a diffractogram: intensity versus diffraction angle $2\theta$. The physically relevant information lies in the position, the power and the half-width of the peaks. For the physical background of x-ray diffractometry of thin films as well as the interpretation of the data obtained, we refer to Birkholz (2006). The peak positions are characteristic for the crystalline structures present in the sample. Small shifts of the peaks with respect to the ideal positions are often related to mechanical strain in the crystalline lattice arising from lattice imperfections introduced during thin film preparation. From the peak power the relative abundance of a specific crystalline orientation can be estimated allowing the determination of the texture of crystalline orientations. Such an analysis has been performed, for example, in the case of thin films of $In_2O_3$:Sn prepared by various deposition techniques [Mergel, Thiele and Qiao (2005)].

The half-width of the peak is related to the crystallite size and to inhomogeneous strain within the crystallites. These parameters are strongly influenced by the preparation conditions and determine to a large degree the optical and electrical properties of the thin films.

Methods for the analysis of x-ray diffractograms have been developed mainly in the context of powder diffractometry. Although the underlying physical principles are the same, some care has to be taken in applying techniques from powder diffractogram analysis to the case of thin films. For a general review of the physical background required for analyzing thin film diffractograms as well as a discussion of similarities and differences between powder and thin film diffractometry, we refer to Chapter 3 of Birkholz (2006).

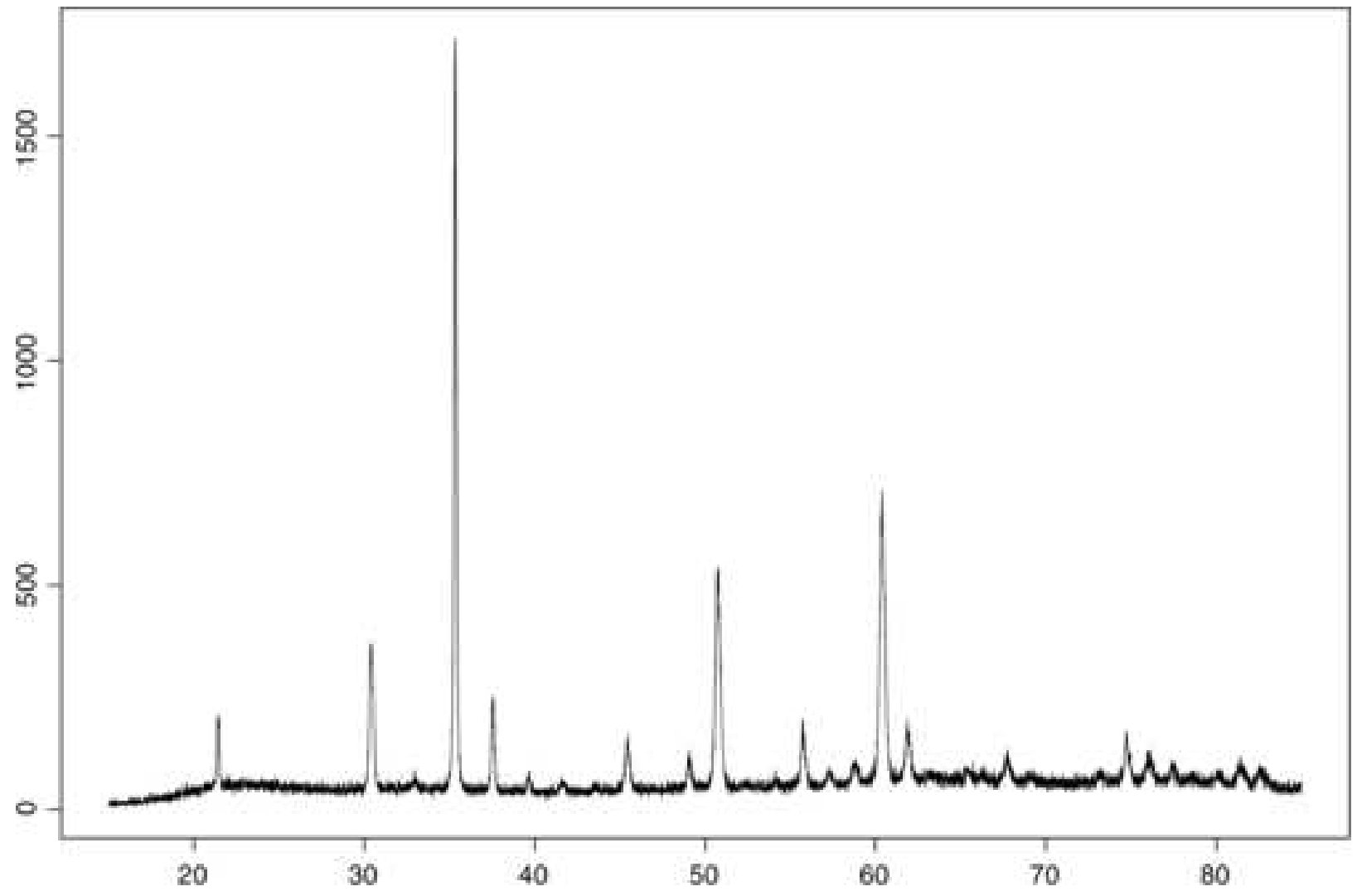

FIG. 1. *The intensity of diffracted x-rays as a function of the angle of diffraction ($2\theta$).*

In our laboratory practice, we have so far used an ad-hoc method to evaluate the x-ray diffractograms. It has proved to be adequate when the potential peak positions were known a priori, that is, in cases where the produced material was already identified [Mergel, Thiele and Qiao (2005)]. With this method, the baseline of the data, arising from the noise level of the signal channel, was taken as a piecewise linear interpolation between the intensity values at positions in the middle between two neighboring theoretical positions. Denoising was done by averaging the data in a pre-defined abscissa interval. The peak position was then looked for in the vicinity of the theoretical positions and the shape of the peak was fitted with a Gaussian kernel. This method uses optimization criteria for noise that are statistically not well founded and shape functions that are often inadequate for x-ray peaks. Furthermore, in the general case, the crystalline structures in the films are not known and, therefore, diffraction peaks can occur at arbitrary values of $2\theta$. This means that the search procedure as a whole is not applicable. The automatic procedure we describe below does not rely on a-priori knowledge of peak positions and uses more flexible models of the baseline, the peaks, the kernels and the noise.

A typical data set is shown in Figure 1. The material under consideration here is a thin film of $In_2O_3$:Sn. Although it is not obvious from the figure, the data, being counts of photons, are integers. A simple stochastic model for the counts at an angle of diffraction $2\theta$ is the Poisson distribution with mean $f(2\theta)$ for an appropriate function $f$. The noise present in the data does not exhibit any obvious dependencies so that a Poisson model is completely specified by fixing $f$ and then taking the observations at each $2\theta$ to be independently distributed. For large $f(2\theta)$ this accurately describes the noise level, but for small $f(2\theta)$ the noise level is underestimated because of ground noise due to the electronics. In practice, the standard deviation of the noise is at least 7 which, in the Poisson model, corresponds to a mean of about 50. For such large parameter values the Poisson distribution can be adequately modeled by a normal distribution with mean and variance 50. This leads to the model

$$Y(t) = f(t) + \sigma(t)Z(t), \qquad 0 \le t \le 1, \tag{2}$$

where $f : [0, 1] \to \mathbb{R}$, $Z(t)$ is standard Gaussian white noise and $\sigma(t)$ is a function of $f(t)$. We have here followed standard practice in statistics and defined the model on the interval $[0, 1]$, whereas the actual data are defined for $2\theta$ in $[15, 85]$. This should cause no problems. In thin film diffractometry measurements are usually taken for equidistant angles, but the method we propose holds also for nonequispaced measurements.

The procedure we propose determines the number, positions, powers and shapes of the relevant peaks as measured from the baseline and then decomposes them into a sum of kernels. It consists of five steps:

(1) The data are approximated using the taut string method. This yields a first estimate of the number, the positions and the heights of the peaks.
(2) An estimate of the first derivative is obtained from a weighted smoothing spline fitted to the original data.
(3) The peak intervals are determined from the positions of peaks according to (1) and a threshold for the derivative obtained in (2).
(4) The baseline is obtained by fitting a spline to the remaining data set after removing the peak intervals.
(5) The peaks with the baseline subtracted are fitted within their respective intervals by a sum of Pearson Type VII curves.

**3. The approximation or confidence region.** We now describe the construction of the confidence region which provides the basis of our concept of approximation. The following is based on Davies, Kovac and Meise (2008). Suppose we have data $\mathbf{Y}_n = \{(t_i, Y(t_i)), i = 1, \ldots, n\}, 0 \le t_1 < \cdots < t_n \le 1$ which are generated under the model

$$Y(t) = f(t) + \sigma Z(t), \qquad 0 \le t \le 1. \tag{3}$$

This differs from the model (2) only in that we here assume a constant noise level $\sigma$. For any function $g : [0, 1] \to \mathbb{R}$, we define the residuals by

$$r(\mathbf{Y}_n, t_i, g) = Y(t_i) - g(t_i) \tag{4}$$

and the standardized sums of the residuals over intervals $I \subset \{1, \dots, n\}$ by

$$w(\mathbf{Y}_n, I, g) = \frac{1}{\sqrt{|I|}} \sum_{i \in I} r(\mathbf{Y}_n, t_i, g), \tag{5}$$

where $|I|$ denotes the number of points $t_i$ with $i$ in $I$. For a given family $\mathcal{I}_n$ of intervals of $\{1, \dots, n\}$, an $\alpha$-confidence region for $f$ is given by

$$\mathcal{A}_n = \mathcal{A}(\mathbf{Y}_n, \sigma, \mathcal{I}_n, \tau_n) = \left\{ g : \max_{I \in \mathcal{I}_n} |w(\mathbf{Y}_n, I, g)| \leq \sigma \sqrt{\tau_n \log n} \right\}, \tag{6}$$

where $\tau_n = \tau_n(\alpha)$ is chosen such that

$$P\left( \max_{I \in \mathcal{I}_n} \frac{1}{\sqrt{|I|}} \left| \sum_{t_i \in I} Z(t_i) \right| \leq \sqrt{\tau_n \log n} \right) = \alpha. \tag{7}$$

To see this, we note that if the data were generated under (3), then (7) implies that $P(f \in \mathcal{A}_n) = \alpha$. A function $g$ belongs to $\mathcal{A}_n$ if and only if its vector of evaluations at the design points $(g(t_1), \dots, g(t_n))$ belongs to the convex polyhedron in $\mathbb{R}^n$ which is defined by the linear inequalities

$$\frac{1}{\sqrt{|I|}} \left| \sum_{t_i \in I} (Y(t_i) - g(t_i)) \right| \leq \sigma \sqrt{\tau_n \log n}, \qquad I \in \mathcal{I}_n. \tag{8}$$

We mention that by using an appropriate norm [Mildenberger (2008)] $\mathcal{A}_n$ can also be expressed as a ball centered at $\mathbf{Y}_n$. The family $\mathcal{I}_n$ we use will be a dyadic multiresolution scheme as for wavelets. It consists of all single points $[i, i]$, the pairs $[1, 2], [3, 4], \dots,$ the sets of four $[1, 4], [5, 8]$ etc. and including all final intervals whether or not they are of this form. The procedure is therefore not restricted to sample sizes $n$ which are a power of 2. The number of such intervals is at most $2n$ and this collection has proved sufficiently fine for x-ray diffractograms. An exception is the last step of our procedure, where we consider small segments of the data set separately and use the family of all subintervals of such a segment. The use of such a scheme $\mathcal{I}_n$ forces any function $g$ in $\mathcal{A}_n$ to adapt to the data at all resolution levels from single points to the whole interval. Since the noise level $\sigma$ of the data usually is not known in advance, we derive it from the data by using

$$\sigma_n = \text{Median}\{|Y(t_i) - Y(t_{i-1})|, 2 \leq i \leq n-1\} / (\Phi^{-1}(0.75)\sqrt{2}). \tag{9}$$

Now $\mathcal{A}_n = \mathcal{A}(\mathbf{Y}_n, \sigma_n, \mathcal{I}_n, \tau_n)$ is no longer exact, but it is honest [Li (1989)] in that the coverage probability is now at least $\alpha$ [Davies, Kovac and Meise (2008)]. The value of $\tau_n$ in (7) can always be determined by simulations for any $n$ and $\alpha$. It follows, however, from a result of Dümbgen and Spokoiny (2001) on the uniform modulus of continuity of the Brownian motion that $\lim_{n \to \infty} \tau_n = 2$ whatever $\alpha$. A much more precise result is given in Kabluchko (2007). In practice, we use the default value $\tau_n = 2.5$, which has proved satisfactory for the thin film data sets. If it is more important to capture all relevant peaks, even small ones, at the cost of random fluctuations being identified as a peak, a lower value of $\tau_n$ can be chosen and even calibrated to give a desired false discovery rate [Benjamini and Hochberg (1995)].

For data $\mathbf{y}_n = \{(t_i, y(t_i)), i = 1, \dots, n\}$ not necessarily generated the model (3), we refer to $\mathcal{A}(\mathbf{y}_n, \sigma_n, \mathcal{I}_n, \tau_n)$ as an approximation region. Any function $f_n \in \mathcal{A}(\mathbf{y}_n, \sigma_n, \mathcal{I}_n, \tau_n)$ will be regarded as an adequate approximation to the data $\mathbf{y}_n$.

As mentioned in Section 2 for the thin film data, the noise level for large values of $y(t_i)$ is of the order $\sqrt{y(t_i)}$ and is consequently underestimated by $\sigma_n$ of (9). At the same time, for small values of $y(t_i)$, the noise is underestimated by $\sqrt{y(t_i)}$ but correctly estimated by $\sigma_n$. This leads to the model (2) with

$$\sigma(t) = \max(\sigma, \sqrt{f(t)}). \tag{10}$$

We overcome this additional complexity by first obtaining an adequate approximation $f_n$ of the data based on the $\sigma_n$ of (9). We then take the noise level at the angle of diffraction $t_i$ to be

$$\Sigma_n(t_i) = \max(\sigma_n, \sqrt{f_n(t_i)}). \tag{11}$$

We must now replace (4) by

$$\tilde{r}(\mathbf{y}_n, t_i, g, \mathbf{\Sigma}_n) = \frac{y(t_i) - g(t_i)}{\Sigma_n(t_i)} \tag{12}$$

and (5) by

$$\tilde{w}(\mathbf{y}_n, I, g, \mathbf{\Sigma}_n) = \frac{1}{\sqrt{|I|}} \sum_{i \in I} \tilde{r}(\mathbf{y}_n, t_i, g, \mathbf{\Sigma}_n). \tag{13}$$

The resulting confidence region $\tilde{\mathcal{A}}_n$ is then given by

$$\begin{aligned} \tilde{\mathcal{A}}_n &= \tilde{\mathcal{A}}(\mathbf{y}_n, \mathbf{\Sigma}_n, \mathcal{I}_n, \tau_n) \\ &= \left\{ g : \max_{I \in \mathcal{I}_n} |\tilde{w}(\mathbf{y}_n, I, g, \mathbf{\Sigma}_n)| \le \sqrt{\tau_n \log n} \right\}. \end{aligned} \tag{14}$$

The approximation regions $\mathcal{A}_n$ and $\tilde{\mathcal{A}}_n$ include many functions which are of no interest. For example, all functions $g$ which interpolate the data belong to both. Interest always centers on the simplest functions where the definition of simplicity depends on the problem at hand. To detect peaks we are interested in minimizing the number of peaks subject to the function lying in $\mathcal{A}_n$ or $\tilde{\mathcal{A}}_n$. We accomplish this by using the taut string method which is described in the next section. The taut string estimate is a piecewise constant function and is therefore not suitable for identifying the baseline. As the baseline is a slowly varying function, it can be associated with a small first derivative. The second concept of simplicity we use is therefore based on smoothness and is defined by

$$\int_0^1 g^{(2)}(t)^2\,dt. \tag{15}$$

This leads to a problem of quadratic programming which is not feasible for data sets as exhibited in Figure 1. The sample size is $n = 7001$ and the data show a high degree of local variability. We therefore use an approximate procedure based on weighted smoothing splines which results in a cubic spline. We use its first derivative to identify the baseline.

The idea of approximation regions for nonparametric regression is implicit in Davies (1995), where it is based on runs of the signs of the residuals. Both definitions are used explicitly in Davies and Kovac (2001). Approximation regions based on the sums of the signs of the residuals over intervals rather than the residuals themselves have been given by Dümbgen (2003, 2007) and Dümbgen and Johns (2004).

**4. The taut string method.** In this section we give a short description of the taut string method based on a small artificial data set. Panel 1 of Figure 2 shows data generated under (2) with $f(t) = 2.5\sin(4\pi t)$ evaluated at the points $t_i = i/32, i = 1,\ldots,32$ and with $\sigma = 1$. The first step is to calculate the partial sums of the observations $Y(t_i)$:

$$S_Y(t_i) = \frac{1}{n}\sum_{j=1}^{i} Y(t_i), \qquad i = 1,\ldots,n,\ S_Y(0) = 0. \tag{16}$$

These are shown in panel 2 of Figure 2. We now form a tube centered on the cumulative sums with an upper bound $U$ and a lower bound $L$ defined by

$$U(t_i) = S_Y(t_i) + \epsilon, \qquad 1 \le i \le n-1,\ U(0) = 0,\, U(1) = S_Y(1), \tag{17}$$

$$L(t_i) = S_Y(t_i) - \epsilon, \qquad 1 \le i \le n-1,\ L(0) = 0,\, L(1) = S_Y(1). \tag{18}$$

The boundary conditions $U(0) = L(0) = 0$ and $U(1) = L(1) = S_Y(1)$ are chosen to reduce edge effects. The resulting tube is shown in panel 3 of Figure 2. The taut string function $TS$ is best understood by imagining a string constrained to lie within the tube and tied down at $(0,0)$ and $(1, S_Y(1))$, which is then pulled until it is taut (cf. panel 4 of Figure 2). There are several equivalent analytic ways of defining this. The taut string is a linear spline with automatic choice of knots. Panel 5 of Figure 2 shows the knot locations. As an estimate $f_{ts,n}$ of $f$, we take the right derivative of the taut string, except at the last point where we take the left derivative. Closer consideration shows that this can be improved. The derivative of the taut string has a local maximum when the taut string switches from the upper to the lower boundary. The value of the derivative on this section is therefore less than the mean of the $Y$-values. Thus, if we define the estimate at crossover intervals as the mean of the $Y$-values between the knots, we obtain a better approximation without altering the number of local extremes. The same reasoning applies to local minima. The function $f_{ts,n}$ obtained in this manner is shown in panel 6 of Figure 2. The connection with the number of local extremes which explains the efficacy of the method is the following. Consider all absolutely continuous functions $H$ which are constrained to lie within the tube. Then the derivative of the taut string $f_{ts,n}$ has the smallest number of local extreme values and, in particular, it has the smallest number of peaks.

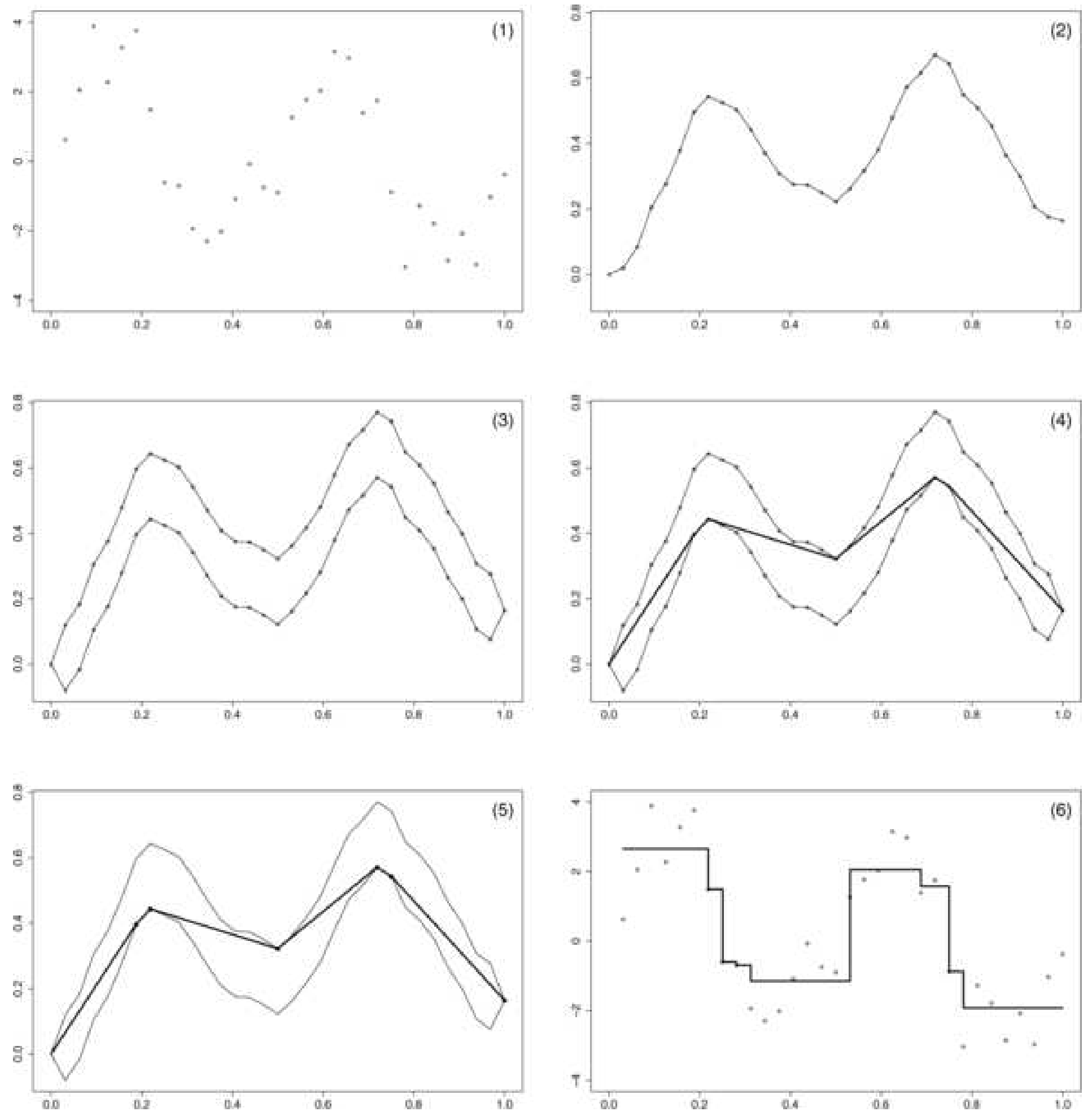

FIG. 2. *Panel (1) shows some noisy sine data. Panel (2) shows the cumulative sums of the data. Panel (3) shows the tube derived from the cumulative sums. Panel (4) shows the taut string through the tube. Panel (5) shows the taut string through the tube with marked knots. Panel (6) shows the data with the right-hand derivative of the taut string.*

This still leaves open the question of the diameter of the tube. Since $\epsilon$ controls the closeness to the data, the basic idea is to start with a very large $\epsilon$ which contains the integral of the mean of the data which, in this case, is the taut string solution $f^1_{ts,n}$. Using the confidence region $\mathcal{A}_n$, we determine those intervals $I \in \mathcal{I}_n$ for which $|w(\mathbf{y}_n, I, f^1_{ts,n})| > \sigma_n\sqrt{\tau_n \log n}$. For any point $i$ which lies in any such interval, we reduce the diameter of the tube by a fixed factor $q < 1$ at the points $t_i$ and $t_{i+1}$. The default value of $q$ which we use is 0.9. A new taut string estimate $f^2_{ts,n}$ is calculated and the procedure repeated in the obvious manner until the estimate lies in $\mathcal{A}_n$. The default version of the taut string uses $\sigma_n$ as specified by (9), $\tau_n = 2.5$ and $\mathcal{I}_n$ as the dyadic multiresolution scheme defined above. The method is fully automatic and does not require the choice of a tuning parameter. As $\mathcal{I}_n$ contains at most $2n$ intervals and the taut string has an algorithmic complexity of $\mathrm{O}(n)$, it follows that the whole procedure has an algorithmic complexity of order $\mathrm{O}(n \log n)$ when the squeezing of the tube is taken into account. Large data sets with $n = 10^6$ and more can be processed in less than one minute.

Panel 1 of Figure 3 shows the result of applying this procedure to an x-ray diffractogram. As mentioned above, this underestimates the noise level for large values of $y(t)$ and may result in side lobes on the large peaks. One such peak is shown in panel 2 of Figure 3. We denote this initial estimate by $\tilde{f}_{ts,n}$ and use it in the definition of $\mathbf{\Sigma}_n$ of (11). This gives rise to the approximation region $\tilde{\mathcal{A}}_n$ of (14) and we can now repeat the taut string procedure. The result is denoted by $f^\star_{ts,n}$. Figure 4 shows $f^\star_{ts,n}$ for the data set of Figure 1. As it can be seen in panel 3 of Figure 3, which shows $f^\star_{ts,n}$ of the same section as $\tilde{f}_{ts,n}$ in panel 2, the side lobes have been removed while leaving the rest of the initial estimate unaltered. It is clear that the automatic taut string method as just described has produced very good resolutions of the peaks and it has not created peaks where none should be.

**5. Heteroscedascity of the ground noise.** Although the ground noise due to the electronics is usually more or less constant, there are cases where it varies sufficiently to cause the taut string to generate additional peaks. This can be avoided by making a nonparametric estimate of the noise level. We follow the approach of Davies (2006). Consider the model

$$V(t) = \sigma(t)Z(t), \qquad 0 \le t \le 1, \tag{19}$$

where $\sigma(t) > 0$ for all $t$ and $Z(t)$ is standard white Gaussian noise. Given measurements $\mathbf{V}_n = \{(t_i, V(t_i)), i = 1, \ldots, n\}$, $0 \le t_1 < \cdots < t_n \le 1$ generated

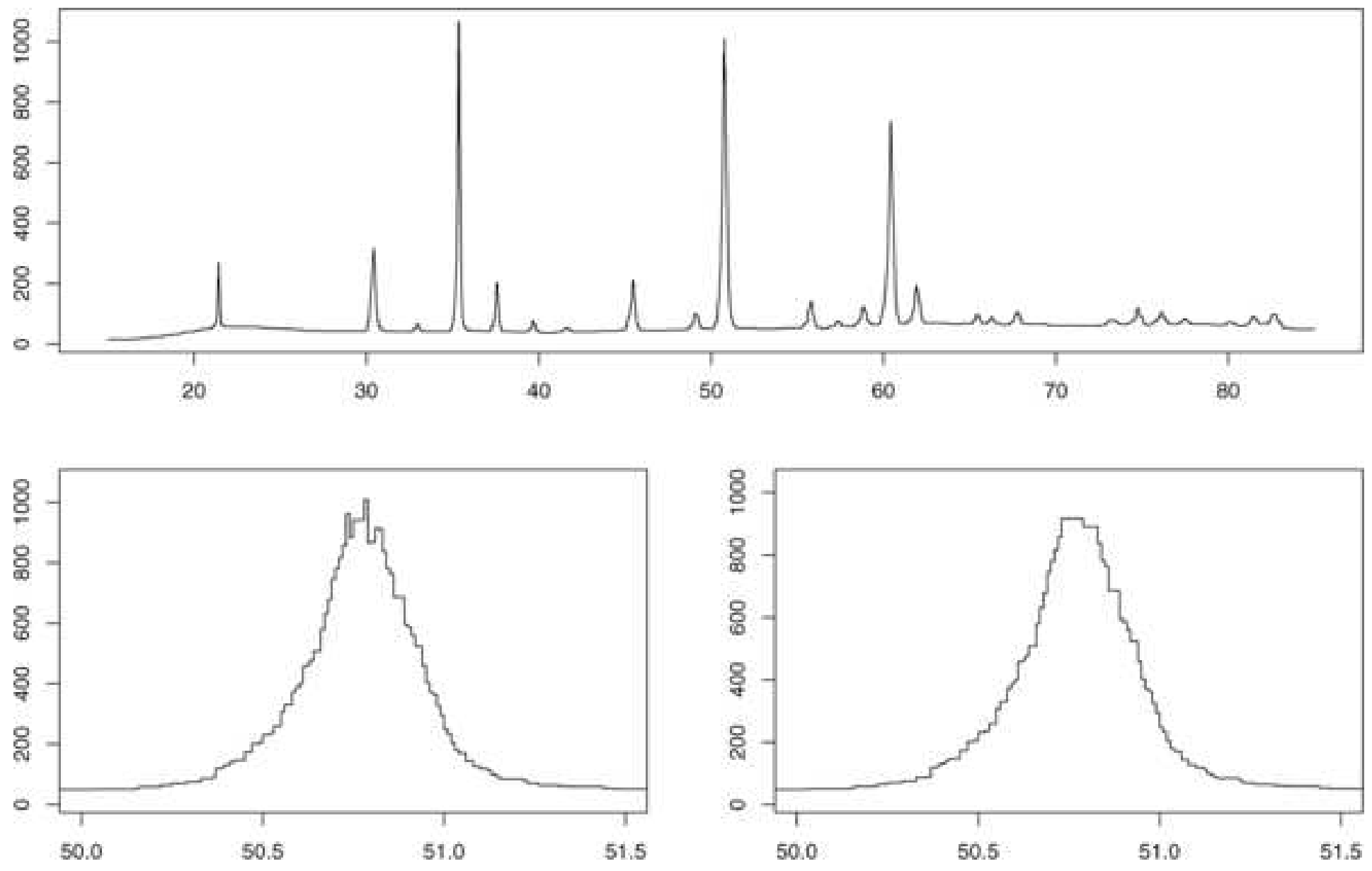

FIG. 3. *Top: One de-noised x-ray diffractogram, using the constant noise estimate given in (9) and a section (bottom left). Bottom right: The same section for the approximation using the local noise estimate given in (11).*

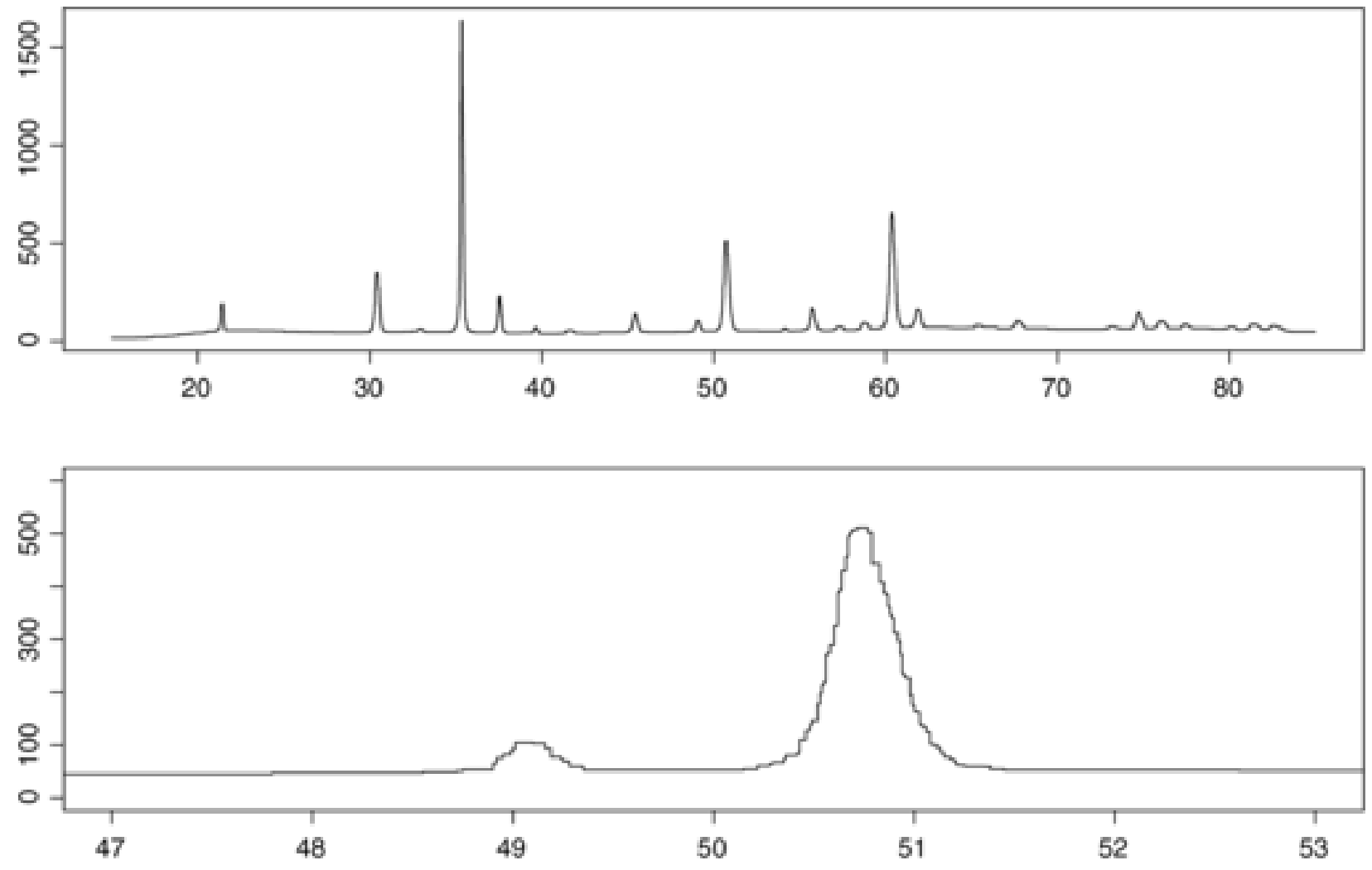

FIG. 4. *The de-noised data of Figure 1.*

under (19), it follows that, for any interval $I \subset \{1, \ldots, n\}$,

$$\sum_{i \in I} V(t_i)^2 / \sigma(t_i)^2 \stackrel{D}{=} \chi^2_{|I|}, \tag{20}$$

where, as before, $|I|$ denote the number of points $t_i$ with $i \in I$. We write

$$\begin{aligned} \mathcal{A}_n^\sigma &= \mathcal{A}_n^\sigma(\mathbf{V}_n, \mathcal{I}_n, \alpha_n) \\ &= \Bigg\{ s : \mathrm{qchisq}(1-\alpha_n, |I|) \le \sum_{i \in I} V(t_i)^2 / s(t_i)^2 \le \mathrm{qchisq}(\alpha_n, |I|) \\ &\qquad \text{for all } I \in \mathcal{I}_n \Bigg\}, \end{aligned} \tag{21}$$

where (i) $\mathcal{I}_n$ denotes a family of subintervals of $\{1, \dots, n\}$, (ii) $s$ denotes a strictly positive function defined on $[0,1]$ and (iii) $\mathrm{qchisq}(p,k)$ denotes the $p$th quantile of the chi-squared distribution with $k$ degrees of freedom. For any given $\alpha$, we may choose $\alpha_n$ so that $\mathcal{A}_n^\sigma$ is a universal, exact and non-asymptotic confidence region for $\sigma(t)$. The value of $\alpha_n$ can be determined by simulations. More easily, if we take $\mathcal{I}_n$ to be the set of all intervals and define $\alpha_n$ by

$$\alpha_n = 1 - \exp(-0.5\tau \log(n)) / \sqrt{\pi\tau \log(n)} \tag{22}$$

with $\tau = 3$, then it may be checked by simulations that $\mathcal{A}_n^\sigma$ has a coverage probability of at least 0.95 for $500 \le n \le 10000$. The regularization we choose is to take $s_n$ to be piecewise constant and then to minimize the number of intervals of constancy subject to $s_n \in \mathcal{A}_n^\sigma$. This is a difficult optimization problem, but the following simplified procedure works well in practice. We start with the interval $J_1 = [1,1]$ and put

$$s_{J_1}^2 = \frac{1}{|J_1|} \sum_{i \in J_1} V(t_i)^2.$$

We now check the inequalities defining $\mathcal{A}_n^\sigma$, but only for all intervals $I \subset J_1$. These clearly hold for $J_1 = [1,1]$. At the $k$th stage $J_1 = [1,k]$, we increase $J_1$ by including $k+1$ and continue in this manner until we reach the end of the sample or, for some $k$ the conditions do not hold for $J_1 = [1, k+1]$. We then take $[1,k]$ as the first interval of constancy and repeat the same process for the second interval. This procedure is continued in the obvious manner until the last sample point is reached. Although this does not solve the original optimization problem, it is clear that it gives a lower bound to the number of intervals of constancy. Moreover, it is not difficult to show that if $\sigma(t)$ of (19) is piecewise constant on a finite number of nondegenerate intervals, then the procedure will consistently estimate the number of intervals and their endpoints.

To allow for heteroscedascity, we proceed as follows. First we use the taut string method with a constant $\sigma_n$ given by (9) to give a first approximation $f_n$. We calculate the residuals and approximate their absolute values by a

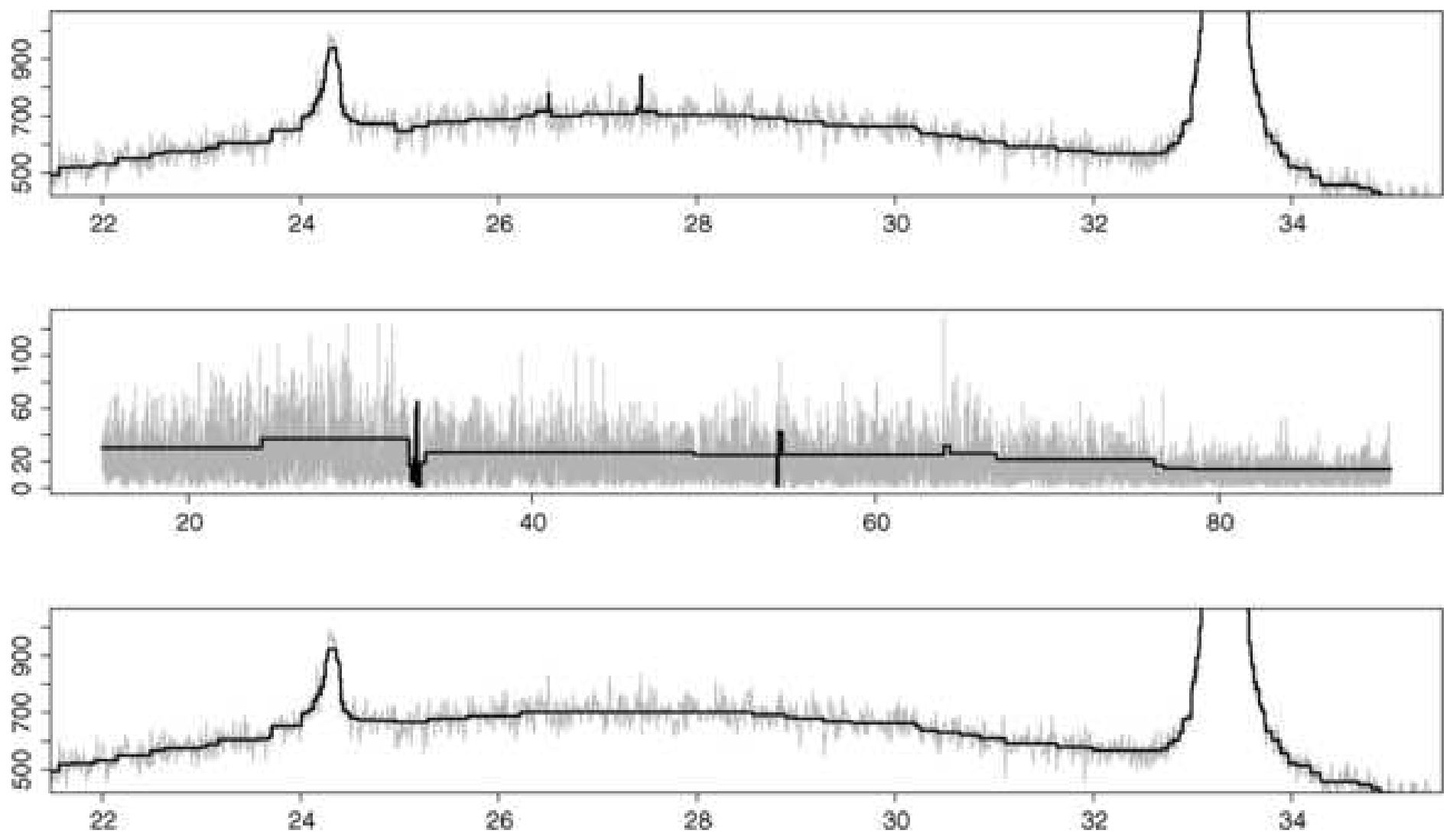

FIG. 5. *The upper panel shows the taut string reconstruction with constant ground noise. The center panel shows the absolute values of all the residuals together with the piecewise constant approximation. The bottom panel shows the final reconstruction for the same section as the top panel.*

piecewise constant function $s_n(t)$ as described above. Finally we replace (11) by

$$\Sigma_n(t_i) = \max(s_n(t_i), \sqrt{f_n(t_i)}) \tag{23}$$

and then proceed as before. The upper panel of Figure 5 shows a section of a data set and the taut string approximation with additional peaks between the angles 26 and 30. The center panel shows the absolute values of all residuals together with the piecewise constant approximation $s_n(t)$: the heteroscedascity is evident. The small values of $s_n(t)$ are due to two large peaks where the residuals are very small. The bottom panel shows the taut string reconstruction using (23).

**6. Weighted smoothing splines.** After having determined the number and locations of the peaks, the next step is to identify the baseline. We do this by fitting a smooth function to the data and then identifying the baseline by the size of the first derivative. As mentioned above, ideally we would like to minimize (15) subject to $g \in \tilde{\mathcal{A}}_n$, using $f^{\star}_{ts,n}$ in (11) or in the heteroscedastic case (23). As $\tilde{\mathcal{A}}_n$ is defined by a series of linear inequalities this, after discretization, leads to a quadratic programming problem which is in principle solvable, but for large data sets and/or data with large variations in local smoothness there are considerable numerical problems. Because of this we take an approach based on weighted smoothing splines which is as follows [Davies and Meise (2008)]. For given weights $\lambda = (\lambda_1, \dots, \lambda_n)$, we consider the solution of the following minimization problem:

$$S_\lambda(g) := \sum_{i=1}^{n} \lambda_i (Y(t_i) - g(t_i))^2 + \int_0^1 (g^{(2)}(t))^2 \, dt \longrightarrow \min! \tag{24}$$

The solution is a natural cubic spline which we denote by $f_{wss,n}$. The weights $\lambda$ are data dependent and chosen to ensure that $f_{wss,n} \in \tilde{\mathcal{A}}_n$. As the smoothness of the solution of (24) increases when the values of the $\lambda_i$ decrease, we wish to choose the weights $\lambda$ to be as small as possible subject to $f_{wss,n} \in \tilde{\mathcal{A}}_n$. We do this in a manner similar to that used in the taut string procedure. We start with very small weights $\lambda_i$ so that the solution is almost a straight line which we denote by $f^1_{wss,n}$. We determine those points $t_i$ which lie in intervals $I$ for which $|\tilde{w}(\mathbf{y}_n, I, f^1_{wss,n}, \mathbf{\Sigma}_n)| > \sqrt{\tau_n \log n}$. At such points we increase the $\lambda_i$ by a factor of $q > 1$. The default value we use is $q = 2$. The solution $f^2_{wss,n}$ is calculated and the procedure is continued in the obvious manner until the solution lies in $\tilde{\mathcal{A}}_n$. The first panel of Figure 6 shows the result of the weighted smoothing spline, $f_{wss,n}$, for the data set of Figure 1. The second panel shows the first derivative $f^{(1)}_{wss,n}$. The smoothness of the solution can be seen from the third panel of Figure 6 which shows the same section of the data as Figure 4.

**7. Identifying the baseline.** To identify the baseline, we combine the results of the taut string, $f^\star_{ts,n}$, and the weighted smoothing spline approximation, $f_{wss,n}$. The baseline is identified by the size of the derivative $f^{(1)}_{wss,n}$ of $f_{wss,n}$. The taut string estimate is piecewise constant, so first we identify those intervals which correspond to the local maxima of $f^\star_{ts,n}$. For each specified interval we find $t_0$ with $f^{(1)}_{wss,n}(t_0) \approx 0$ and $t_0$ inside or close to the actual interval. Afterward we determine $t_{l_2} \le t_{l_1} \le t_0 \le t_{r_1} \le t_{r_2}$ with

$$|f^{(1)}_{wss,n}(t_{l_i})| \approx |f^{(1)}_{wss,n}(t_{r_i})| \approx \mathrm{Median}(|f^{(1)}_{wss,n}|), \qquad \text{for } i = 1, 2, \tag{25}$$

and $f^{(1)}_{wss,n}(t) \ge 0$ for $t \in [t_{l_2}, t_{l_1}]$ and $f^{(1)}_{wss,n}(t) \le 0$ for $t \in [t_{r_1}, t_{r_2}]$. The initial interval is then extended to $[t_{l_2}, t_{r_2}]$. The final intervals are taken as delimiting the peak. The maximum width of the peaks is bounded by five degrees $2\theta$. The remaining data set is approximated again using a weighted smoothing spline and the result $f_{bl,n}$ is the estimate of the baseline. The upper panel of Figure 7 shows the dataset of Figure 1 with automatically fitted baseline. The lower panel shows the data after the baseline has been subtracted.

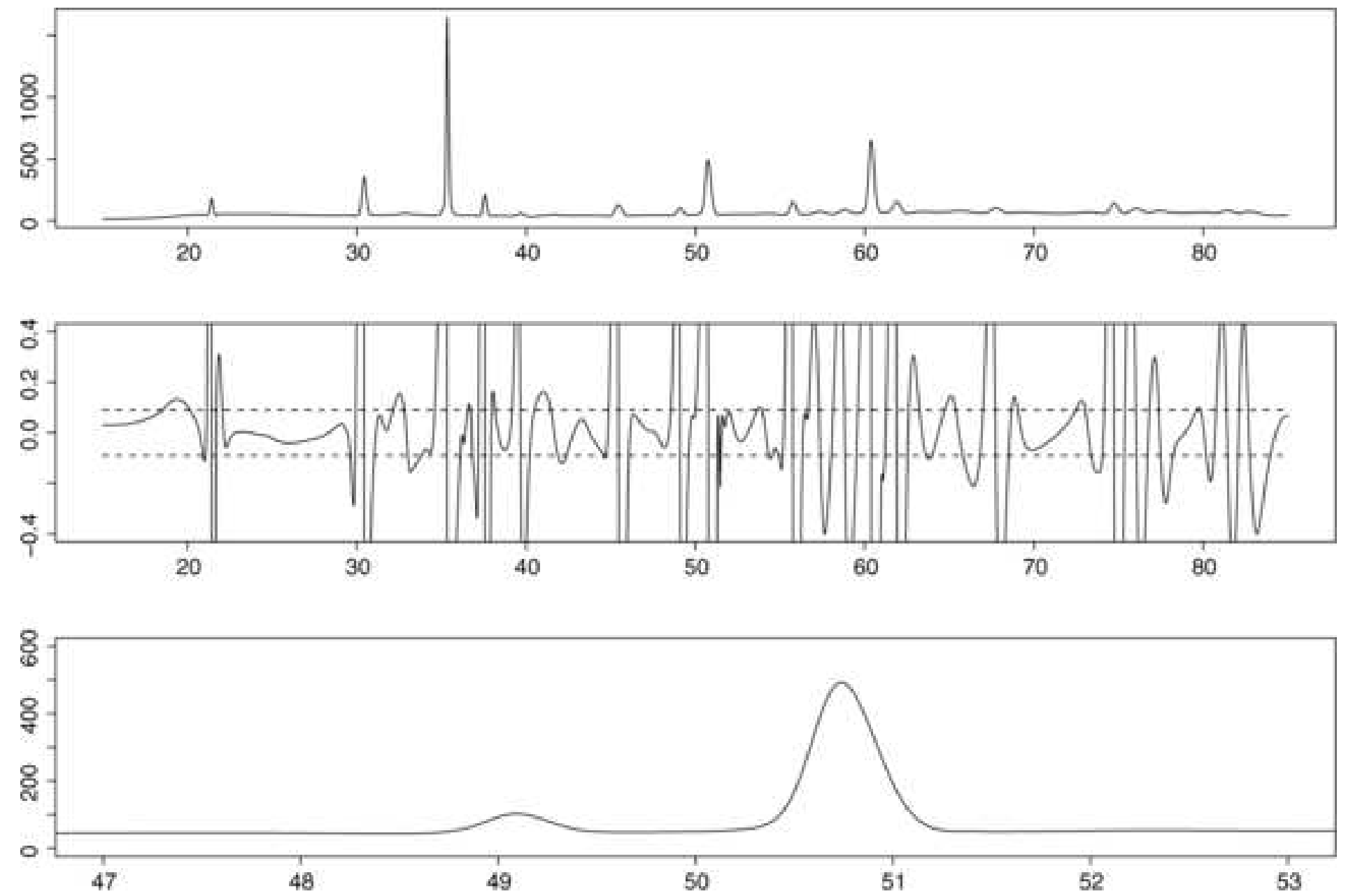

Fig. 6. *The upper panel shows the weighted smoothing spline estimate $f_{wss}$ of the data of Figure 1. The middle panel shows the first derivative $f_{wss}^{(1)}$ together with the used threshold (dotted line) and the bottom panel shows a section of $f_{wss}$.*

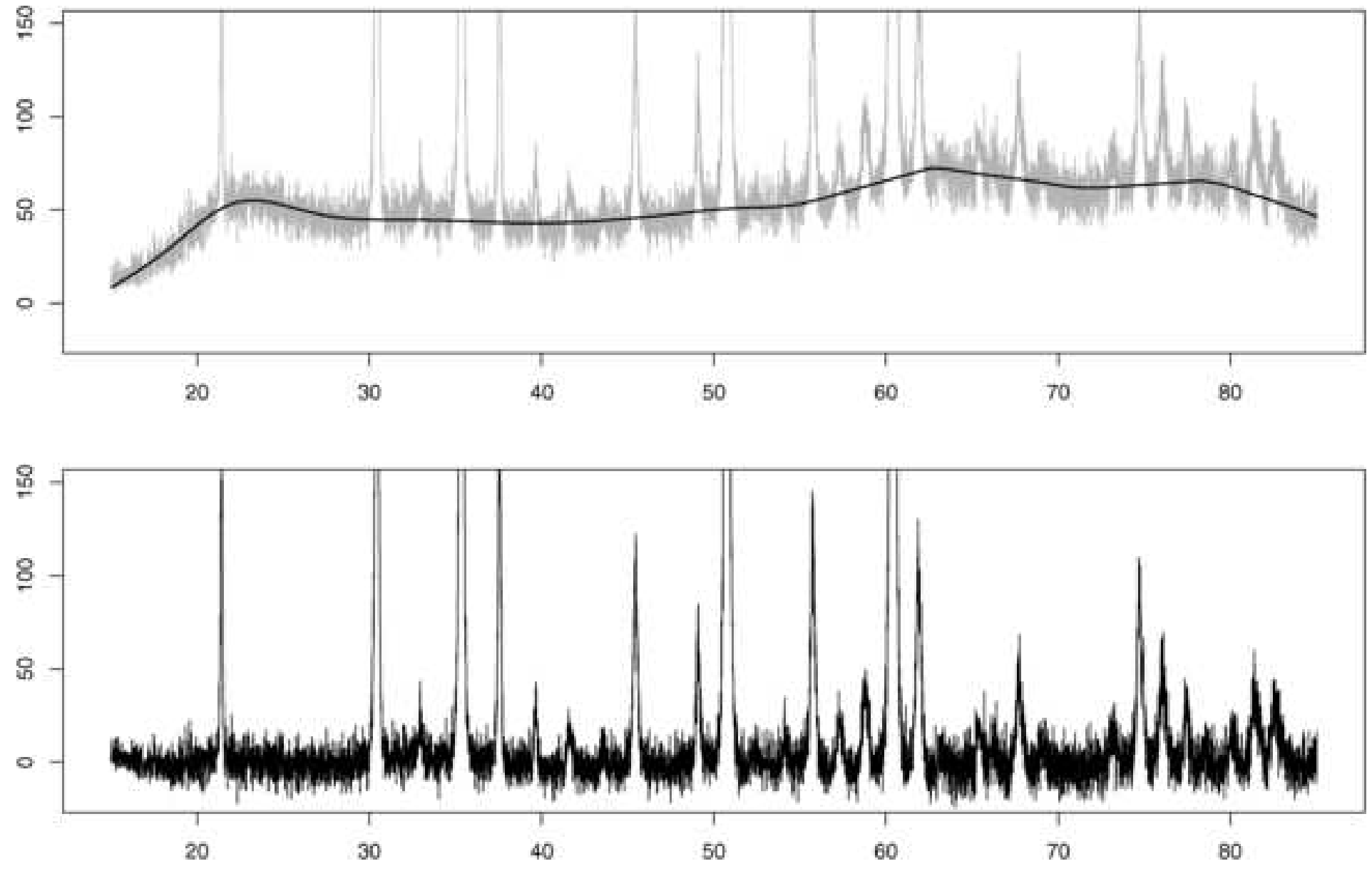

Fig. 7. *Baseline approximation and data with removed baseline.*

**8. The decomposition of the peaks.** We now address the third problem, which is to decompose each peak into a finite sum of kernels. In the simplest case where the kernels are translations of one specified kernel this is a deconvolution problem which is known to be an ill-posed inverse problem. The problem we wish to solve is even more difficult, as the kernels are not fixed but only taken to belong to a parametric family of kernels. Apart from location, they may also differ in width and shape. To solve ill-posed inverse problems, some form of regularization is required. We regularize by minimizing the number of components subject to the condition that the result is consistent with the data in the sense defined in Section 3. In general, the solution will not be unique and stable, as there may be several solutions with very different kernels. We give an example in the next section. The exact mathematical formulation leads to a nonconvex minimization problem with a large number of local minima, so there is no method which is always guaranteed to yield at least one solution with the minimum number of components. We use a random search algorithm which has the advantage of producing different solutions rather than just one. As not all solutions are physically relevant, a deterministic method may well produce a solution, but one which is not physically relevant. An example of a peak with two essentially different solutions is shown in Figure 10 in the next section: No adequate approximation with one kernel is found, but there exist different combinations of two kernels which give adequate approximations to the data.

The algorithm is as follows. First the intervals defined by (25) are treated separately. Let $\{t_l, t_{l+1}, \ldots, t_m\} \subseteq \{t_1, \ldots, t_n\} \subseteq [0,1]$ be the segment under consideration and $L := m - l + 1$ its length. We denote by

$$\tilde{y}(t_i) = y(t_i) - f_{bl,n}(t_i) \qquad \text{for } i = l, \ldots, m$$

the measurements in the interval, where the baseline $f_{bl,n}$ has been subtracted but will subsequently be used for the standardization of residuals. We now construct an approximation to the data $\tilde{y}(t_l), \ldots, \tilde{y}(t_m)$ which we will denote by $f_{pk,n}(t)$. Note that $f_{pk,n}$ is only defined on the peak intervals.

In much the same manner as in the previous sections, we start with the simplest model, one kernel, and then check whether an adequate approximation $f_{pk,n}(t_l), \ldots, f_{pk,n}(t_m)$ to the data exists, that is, whether the appropriately standardized residuals satisfy

$$|\tilde{w}(\tilde{\mathbf{y}}, I, f_{pk,n}, \tilde{\mathbf{\Sigma}}_{\mathbf{n}})| \le C_L \tag{26}$$

for all intervals $I \subseteq \{t_l, \ldots, t_m\}$. We give details on the choice of the set of intervals, the noise level $\tilde{\mathbf{\Sigma}}_{\mathbf{n}}$ and the threshold $C_L$ below, after the description of the procedure. Model complexity (the number of kernels) is increased until the criterion is satisfied. Physical characteristics of interest like power, full width at half maximum (FWHM) and exact location of the peak components can then be calculated from the estimated components.

Each decomposition is of the form

$$f(t) = \sum_{i=1}^{k} \gamma_i p(t; \beta_i), \tag{27}$$

where $k$ denotes the number of kernels (starting with $k = 1$) and $\gamma_i$ are nonnegative weights. The kernels $p$ depend on a vector of parameters $\beta_i$ including location and shape parameters. Depending on the parameterization, the weights $\gamma_i$ correspond either to the maximum height or to the power of the peak component. The number and interpretation of the parameters as well as the range of admissible values depend on the family of curves used. Several choices of kernels are possible, but the most widely used families all include densities of the Gaussian and Cauchy (Lorentz) distributions as extreme cases, say [Birkholz (2006), Chapter 3]. Among these families are Voigt functions, which are convolutions of Gaussian and Cauchy densities, so-called pseudo-Voigt functions, which are convex combinations of Gaussian and Cauchy densities, and Pearson Type VII curves. The approach presented here is not limited to these families of curves and should work for any suitably chosen parametric family of kernels including asymmetric ones.

In the following, we will only consider the Pearson Type VII family, since it works well for our data and avoids some numerical difficulties that occur when using Voigt or pseudo-Voigt functions. The Pearson Type VII family consists of affine transformations of t-distributions; see Chapter 28.6 in Johnson, Kotz and Balakrishnan (1995). This family has been used for fitting diffraction peaks for a long time; cf. Hall et al. (1977).

The curves have the form

$$p(t; \beta) = p(t; \mu_i, m_i, a_i) = \left(1 + \frac{(t - \mu_i)^2}{a_i^2 m_i}\right)^{-m_i}, \tag{28}$$

where $\mu_i$ is the location parameter, $a_i$ measures the width, and $m_i \geq 1$ determines the shape of the curve. For $m_i = 1$, $p$ is the Cauchy density, and since $(1 + \frac{x^2}{m})^{-m} \stackrel{m\to\infty}{\longrightarrow} \exp(-x^2)$, the shape becomes finally Gaussian for large $m_i$. The kernel is not normalized, so it is not necessarily a probability density. We have $p(\mu_i; \mu_i, m_i, a_i) = 1$, so the weight $\gamma_i$ is the height at the maximum. For each Pearson VII kernel, we have to estimate four parameters: $\mu_i$, $m_i$, $a_i$ and the weight $\gamma_i$.

For fixed $k$ (starting with $k = 1$), we consider signals of the form

$$f_{pk,n}(t) = \beta_0 + \beta_1 t + \sum_{i=1}^{k} \gamma_i p(t; \mu_i, m_i, a_i), \tag{29}$$

where $p(t;\mu_i,m_i,a_i)$ is the Pearson VII function with parameters as described above. The parameters $\beta_0$ and $\beta_1$ are added to allow small changes in the baseline estimate and should only have small values, for examples, values between $\pm d_0 := \pm 5\%$ of the height of the initial baseline estimate for $\beta_0$ and between $-d_1$ and $d_1$ for $\beta_1$. We choose $d_1$ so that the slope of the baseline estimate can change by at most 5 counts per $2\theta$. Thus, the parameter vector

$$\boldsymbol{\beta} = (\beta_0, \beta_1, \gamma_1, \mu_1, m_1, a_1, \dots, \gamma_k, \mu_k, m_k, a_k)$$

completely determines the shape. Since it is not possible to check directly whether an adequate approximation of given complexity $k$ exists which satisfies our criterion, we focus on one or several promising candidates. As the estimate is required to be close to the data with small residuals, a natural choice is the nonlinear weighted least squares estimate, where the weights are the reciprocals of the scale estimates (11). In the heteroscedastic case, we use (23) instead. This leads to the following optimization problem:

$$R(\boldsymbol{\beta}) = \sum_{j=l}^{m} \left( \frac{f_{pk,n}(t_j;\boldsymbol{\beta}) - \tilde{y}(t_j)}{\Sigma_n(t_j)} \right)^2 \longrightarrow \min!, \tag{30}$$

with

$$f_{pk,n}(t;\boldsymbol{\beta}) = \beta_0 + \beta_1 t + \sum_{i=1}^{k} \gamma_i p(t; m_i, \mu_i, a_i)$$

subject to

$$\begin{aligned} &-d_j < \beta_j < d_j && (j = 0, 1), \\ &\gamma_i, a_i > 0 && (i = 1, \dots, k), \\ &t_l < \mu_1 < \dots < \mu_k < t_m && (i = 1, \dots, k), \\ &m_i \geq 1 && (i = 1, \dots, k). \end{aligned}$$

Simple re-parameterizations can be used to eliminate the interval constraints, for example, logarithms and affine transformations of the logit-function. For $k > 1$ every signal has $k!$ different parameterizations because of interchangeability of the kernels, and a reduction of the search space is achieved by enforcing an ordering in the location parameters $\mu_1 < \dots < \mu_k$. An appropriate transformation is given by Jupp (1978).

Since (30) generally has a large number of local minima, we proceed iteratively in the following manner. We choose a starting value at random from a uniform distribution over a suitably chosen rectangular set which contains all reasonable parameter values. This is followed by a Newton-type procedure to find the nearest local minimum of $R$. We use the so-called BFGS-Method as described in Chapter 3.2 of Fletcher (2000). The local minimum of $R$ is then compared to the lowest value previously found. If it is lower, we check the conditions (26), and stop if they are fulfilled. In this case, an adequate approximation with given complexity has been found. Otherwise, we draw a new starting value at random and repeat these steps. If no adequate approximation is found within a specified number of iterations, the output is the best local minimum of $R$ that has been found. The number of kernels is then increased by one, and the procedure is started anew.

Note that this optimization algorithm does not directly aim at obtaining an adequate approximation in the sense of the criterion (26) but tries to find local optima of the weighted least squares residual function. This is less difficult as $R$ is infinitely differentiable.

When checking (26), we standardize the residuals using

$$\tilde{\mathbf{\Sigma}}_{\mathbf{n}}(t) = \sqrt{f_{bl,n}(t) + f_{pk,n}(t)}$$

instead of (11). In the case of heteroscedastic ground noise, we use the estimate given in (23). Since we only consider one segment of the data at a time, we have a much smaller number of observations. This allows us to use all subintervals of $t_l, \ldots, t_m$ in (26). We use an efficient algorithm given by Bernholt and Hofmeister (2006) for this. The value of $C_L$ is determined by means of simulation, since the asymptotic choice given in Section 3 is not valid for small sample sizes.

It is in the nature of this problem that multiple solutions may exist, especially when fitting two or more kernels. If only one solution is required, then the process could be terminated here. However, in the context of thin film data, some solutions may be physically relevant and others not. For this reason, even after a solution has been found, the process is repeated a fixed number of times set by the user. This provides some idea of the variability of possible solutions for this particular segment of the data. In some cases these will be very similar, but they might also differ strongly. An example of this is described in the next section. The experimenter may then either choose the solution that is the most meaningful, based on partial prior knowledge about possible components of the material under consideration or on the results for the other peak intervals, or decide that no physically meaningful, unambiguous interpretation of this part of the data is possible.

Once a solution is found, the characteristics of the peak components can be estimated by calculating the values for the fitted curves. For Pearson VII curves as used here, the corresponding weight parameter $\gamma_i$ equals the maximum height. The integrated intensity $I_i$ of the $i$th component is obtained by

$$I_i = \frac{\Gamma(m_i - 1/2)\sqrt{\pi m_i} a_i}{\Gamma(m_i)} \gamma_i;$$

cf. Hall et al. (1977). The full width at half maximum of the $i$th kernel depends only on the shape and scale parameters $m_i$ and $a_i$, and can be calculated explicitly by

$$FWHM_i = 2a_i\sqrt{m_i(\sqrt[m_i]{2} - 1)};$$

cf. Hall et al. (1977). Of course, $I_i$ and $FWHM_i$ must be scaled appropriately according to the grid width. If an interval contains two or more strongly overlapping peak components, or if the components have very low intensities, the values calculated may not be reliable.

**9. Physical interpretation of some results.** The diffractogram presented in Figure 1 was taken on a thin film of $In_2O_3$:Sn, that is, indium oxide doped with tin. This material is usually called ITO (indium-tin-oxide). It has a good electrical conductivity and is transparent in the visible wave length region. Its main application is as top electrode in liquid crystal displays (LCDs) or in antireflective coatings on TV monitors [Mergel (2006)]. X-ray diffractograms of such films contain important information on the material's quality, as it was shown that the optical and electrical properties are correlated with the lattice expansion that can be deduced from the position of the peaks [Mergel and Qiao (2004)].

The thin film consists of many small crystalline grains with different crystallographic orientations and the experimental conditions for taking the diffractograms are those of powder diffraction [Smart and Moore (2001)]. X-ray reflexes are observed when a set of grains satisfies the Bragg condition for a particular diffraction angle

$$2d\sin(\theta) = \lambda, \tag{31}$$

where $d$ is the distance between the lattice planes of the grains parallel to the sample surface, $\theta$ is the diffraction angle and $\lambda$ is the wave length of the x-ray radiation used in the diffraction apparatus. To give an example, the peaks in the diffractogram of Figure 1 at about 30.4° and 35.4° are assigned to ITO grains with orientation (222) and (400), respectively. The numbers in parentheses designate the Miller indices.

The Miller indices are characteristic of the orientation of the specimen. For crystals with a cubic lattice, the Miller indices give the vector normal to the planes, (2,2,2) and (4,0,0) for the examples given above and the distance between the planes is obtained from

$$d_{hkl} = \frac{a_0}{\sqrt{h^2 + k^2 + l^2}}, \tag{32}$$

where $a_0$ is the lattice constant, that is, the lateral extension of the unit cell of the crystal and $h$, $k$, $l$ are the Miller indices, $(hkl)$.

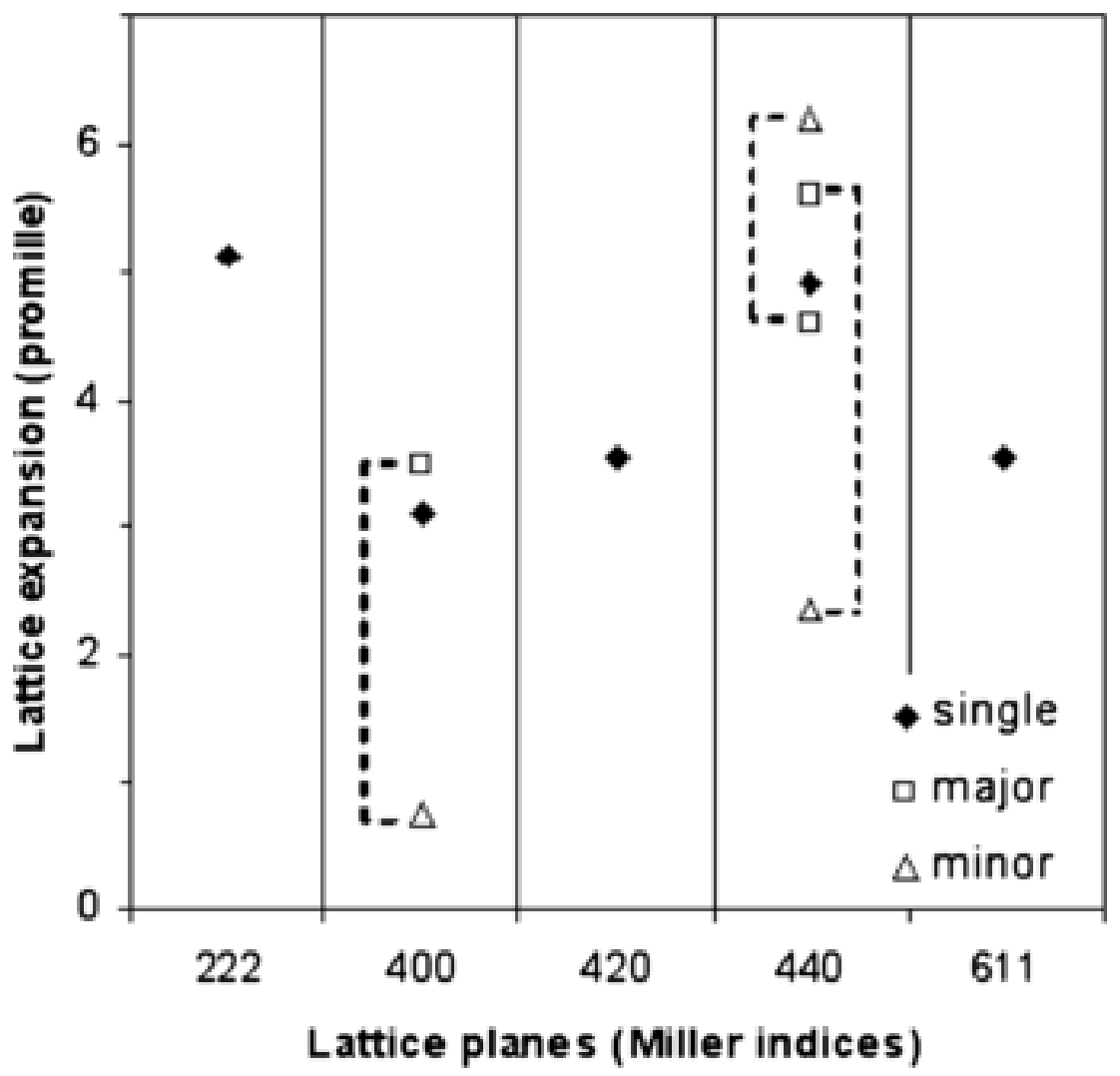

FIG. 8. *Lattice distortion of the peaks represented in Table 1. Only a lattice expansion is observed. The full symbols designate the results of peak fits with one center and the open symbols those where two centers have been used.*

All possible values of $d_{hkl}$ of an ideal cubic crystal can be calculated from equation (32) when the lattice constant is known. In the case of $In_2O_3$, the basic substance of ITO, $a_0 = 1.0118$ nm. From $d_{hkl}$-values the positions of all diffraction peaks can be calculated using the Bragg condition (31). Vice versa, the values for $d$ can be deduced from the position of the diffraction peaks and assigned to $(hkl)$ planes. This has been done for the peaks of Figure 8.

We do not expect ideal crystals in our sample because the thin film has been deposited under nonequilibrium conditions and has been subjected to ion bombardment during film growth. Therefore, it is interesting to evaluate the deviation of the lattice constants of the grains from the ideal value. This deviation is characterized by the lattice distortion $(d-d_0)/d_0 = \Delta d/d_0$, where $d$ denotes the experimentally found value and $d_0$ that of the ideal crystal calculated from equation (32) with the ideal lattice constant. In the case of ITO, the lattice distortion is different for the different grain orientations [Mergel et al. (2000)].

The results of the evaluation of some peaks of the sample are displayed in Table 1 in order to discuss the importance of the statistical analysis for physical problems. The lattice distortion for these peaks is depicted in Figure 8.

The FWHM is correlated with the quality of the crystalline grains. A large FWHM indicates small grains or a large density of lattice defects that are both responsible for the reduction of the correlation length of the crystalline

Table 1

*Evaluation of some peaks of the diffractogram of Figure 1. The first three columns designate the Miller indices (hkl), the peak position and the lattice distortion, respectively. The last column indicates whether the decomposition is accepted by the residual criterion (26)*

| | $2\theta$ | $\Delta d/d_0$[‰] | Height | Intensity | FWHM | $m$ | Accepted |
|---|---|---|---|---|---|---|---|
| (222) | 30.42 | 5.1 | 324 | 96 | 0.27 | 7.2 | Yes |
| (400) | 35.34 | 3.1 | 1548 | 288 | 0.15 | 2.1 | No |
| | 35.33 | 3.5 | 1614 | 250 | 0.12 | 1.5 | |
| | 35.43 | 0.8 | 487 | 51 | 0.07 | 1.0 | Yes |
| (420) | 39.66 | 3.5 | 38 | 11 | 0.18 | 1.0 | Yes |
| (440) | 50.75 | 4.9 | 460 | 170 | 0.34 | 6.4 | No |
| | 50.68 | 6.2 | 131 | 33 | 0.16 | 1.0 | |
| | 50.77 | 4.6 | 378 | 144 | 0.35 | 8.5 | Yes |
| | 50.71 | 5.6 | 420 | 129 | 0.27 | 3.2 | |
| | 50.89 | 2.3 | 155 | 62 | 0.27 | 1.1 | Yes |
| (611) | 55.76 | 3.5 | 119 | 39 | 0.29 | 3.5 | Yes |

lattice. The (222)-oriented grains exhibit a big lattice expansion together with a big FWHM. This correlation is physically meaningful because both effects originate from a strong incorporation of additional oxygen into the lattice leading to an expansion of the lattice and a big density of lattice defects [Mergel et al. (2000)]. The fit of the peak mentioned in Table 1 is shown in the first row of Figure 9.

The (400)-oriented grains are characterized by a small lattice expansion and a small FWHM. This is consistent with literature where it is reported that this orientation is most resistant against bombardment with oxygen and, therefore, incorporates less oxygen than the other orientations [Mergel et al. (2000)]. The residual criterion excludes the solution with one single peak but accepts the two-peak solution reported in Table 1 and Figure 8. This comprises a minor contribution with a lattice constant close to the ideal one indicating that there exists a group of crystalline grains that did not incorporate any additional oxygen at all. Such grains may exist in a certain depth of the film where ideal growth conditions have been met. The fit with one kernel as well as the decomposition into two kernels is shown in the second and third rows of Figure 9, respectively.

The situation is more complicated for the (440)-oriented grains as shown in Figure 10. The single-peak solution is discarded by the residual criterion (26). There exist at least two equally acceptable two-peak solutions, one with the major peak at the leading position and the other one with the major peak at the trailing position. The first case yields two different FWHMs (0.35 and 0.16), similar to the two-peak solution for (400), whereas the second case yields two similar FWHMs. A priori, there is no physical reason to prefer one solution over the other and further investigation is needed to decide.

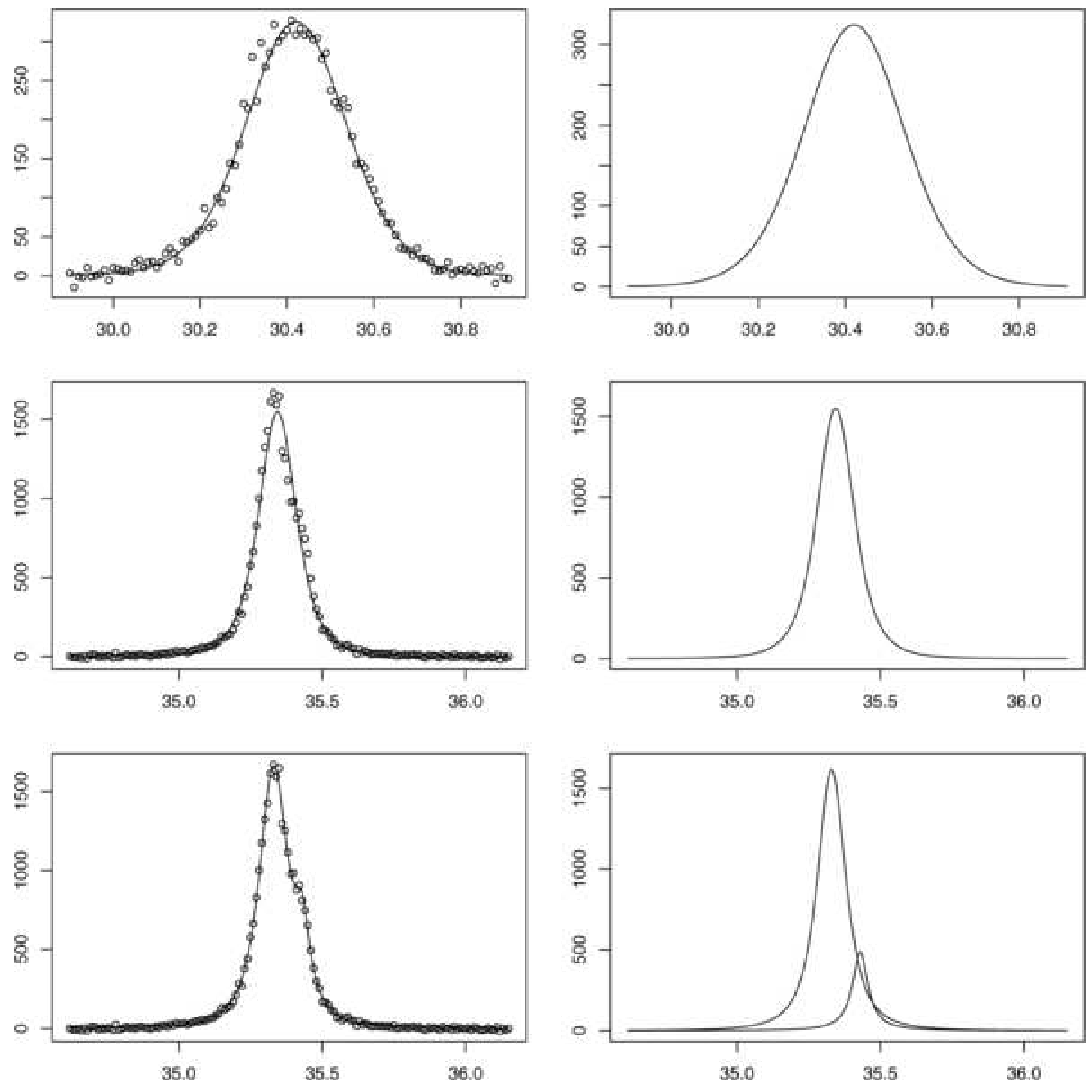

Fig. 9. *Fit for the peaks corresponding to (222)-oriented grains (first row) and fits with one and two kernels for the (400)-oriented grains (second and third rows). The left column shows the data and the resulting fits, the right column shows the individual components.*

The intensity of the peaks is compared to that observed for ideal powder samples. In our case it is found that the (400) grains are much more and (222) grains much less abundant than expected for an equal distribution. This is an indication of a strong ion bombardment during deposition to which (400) planes are not sensitive and which leads to a strong oxygen incorporation in (222) grains hampering their growth.

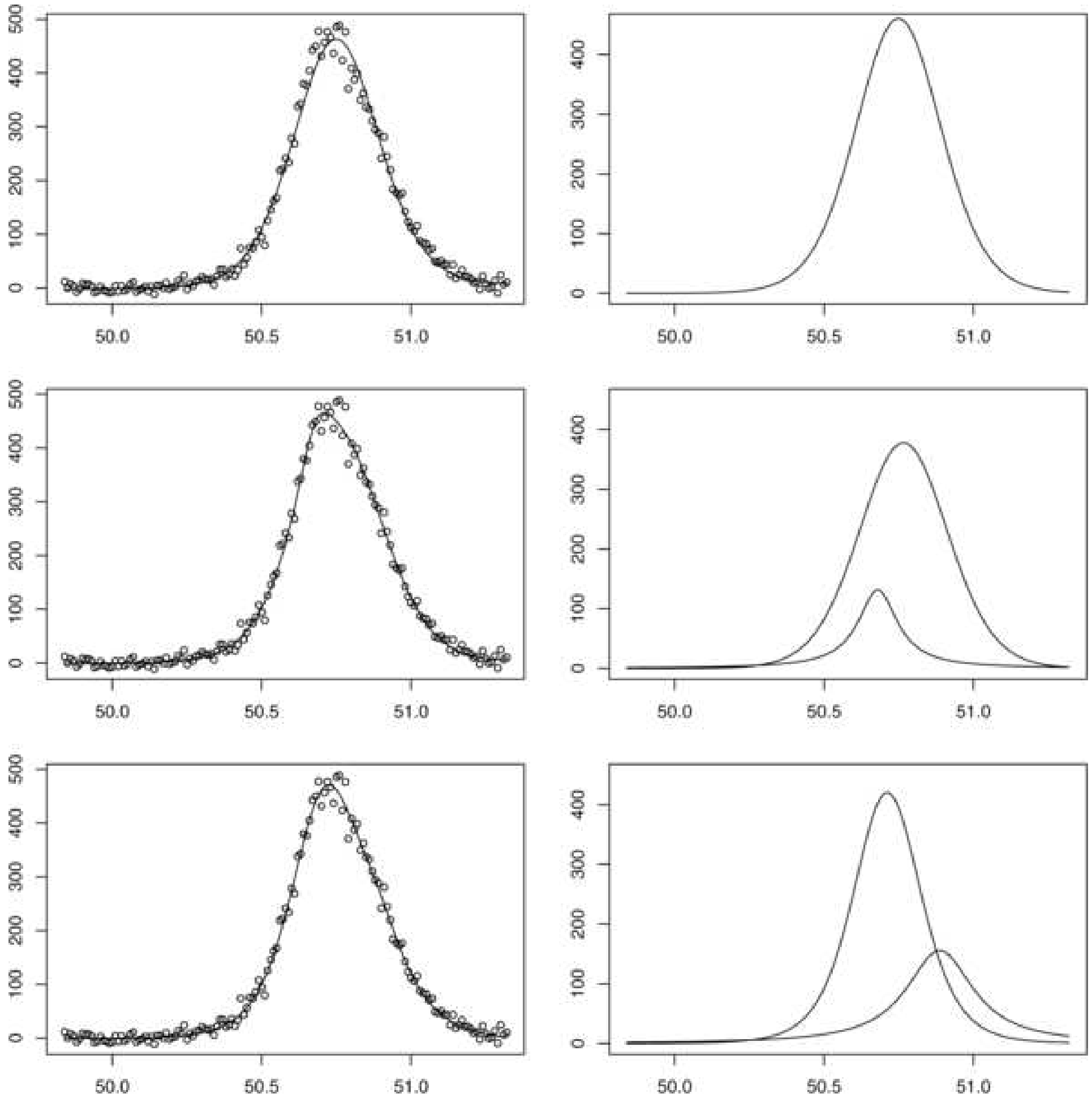

FIG. 10. *Fit for the peaks corresponding to (440)-oriented grains. The fit with one kernel (first row) is rejected, but there exist at least two different fits with two kernels that are accepted by the criterion (second and third rows). The left column shows the data and the resulting fits, the right column shows the individual components.*

**10. Discussion.** In this article we have proposed a fully automatic five-step procedure that determines the number, positions, powers and shapes of the relevant peaks and their components in x-ray diffractograms. It can be applied when little or no prior knowledge of approximate peak positions is available, as is often the case in the analysis of the morphology of thin films. The procedure is based on recent advances in nonparametric regression and denoising techniques, the taut string method and weighted smoothing splines. The taut string method is very successful in producing approximations with a small number of local extremes and is therefore used in step one to determine the positions of the peaks. As the approximation is piecewise constant, it cannot be used for the detection of the boundaries of the peaks, which are necessary for a separate fit of the baseline. This problem is solved using weighted smoothing splines—used in step two—to give an approximation to the data which is twice continuously differentiable. The peak locations derived from the taut string and the first derivative of the fitted spline are used in step three to determine baseline and peak regions of the data set. The threshold of the first derivative used to define the baseline is obviously dependent on the problem under consideration and cannot be decided purely on statistical principles. In step four of the procedure weighted smoothing splines are again used to estimate the baseline which is required to determine the power of the peaks. These first four steps of the procedure are computationally fast and were carried out in about 1 minute on a standard PC (AMD Athlon 64 X2 Dual Core 6000+, 3.01 GHz, 3.21 GB RAM) for the data set used as an illustration throughout the article ($n = 7001$).

The last step of the procedure—decomposition of the peaks—requires the solution of a nonlinear least-squares problem. Here multiple solutions may occur, especially when more than one kernel is fitted. A stochastic search algorithm is used to find several candidate solutions. It is left to the experimenter to decide which of these multiple solutions are physically relevant. Calculation times may vary for this step depending on the number of intervals identified in the previous steps, the number of kernels needed to decompose the peaks and the number of alternative solutions desired. Most of the calculation time is spent attempting to fit a model with a smaller number of components before an upper limit of iterations is reached and the number of components is increased; in some cases this can take several minutes. However, if there is a solution with one kernel, it is usually found within the first few iterations. For the data set presented here, there were 25 intervals, 19 of which consisted of 1-component peaks. Calculation of the peak decompositions took slightly more than 2 minutes for step five, adding up to a total computation time of about 3.5 minutes.

**Software.** The procedure and the data set analyzed in this article are available in the R-package `diffractometry` on CRAN. It can be downloaded from http://www.cran.r-project.org/.

**Acknowledgments.** The authors also wish to thank Malte Gather for helpful discussions and suggestions.

P. L. Davies
M. Meise
University of Duisburg-Essen
Department of Mathematics
45117 Essen
Germany
E-mail: laurie.davies@uni-due.de
monika.meise@uni-due.de

U. Gather
T. Mildenberger
Technische Universität Dortmund
Department of Statistics
44221 Dortmund
Germany
E-mail: gather@statistik.tu-dortmund.de
mildenbe@statistik.tu-dortmund.de

D. Mergel
University of Duisburg-Essen
Department of Physics
45057 Duisburg
Germany
E-mail: dieter.mergel@uni-due.de